\RequirePackage{fix-cm}
\documentclass[smallextended]{svjour3}       
\smartqed  
\usepackage{booktabs}
\usepackage{graphicx}
\usepackage{xcolor}
\usepackage{fancybox}
\usepackage{pdflscape}
\usepackage{caption}
\usepackage{subcaption}
\usepackage{tabularx}
\usepackage{longtable}
\usepackage{array}
\usepackage{fontawesome}
\usepackage{multirow}
\usepackage{array}
\usepackage{easyReview}

\def\mybarhhigh#1#2{
   {\color{black}\rule{#1mm}{4pt}}  #2}
\def\darkgraybar#1#2{
    {\color{darkgray}\rule{#1mm}{4pt}}  #2}
\def\graybar#1#2{
   {\color{lightgray}\rule{#1mm}{4pt}}  #2}

\usepackage{subcaption}
\usepackage{booktabs}
\usepackage{setspace}
\usepackage{tikz}
\usepackage{amssymb}
\usepackage{booktabs}
\usepackage{enumitem}
\usepackage{changepage}
\usepackage{hyperref}
\hypersetup{
    colorlinks=true,
    linkcolor=black,
    filecolor=black,      
    urlcolor=black,
    citecolor=black,
}
\usepackage{graphicx}

\usepackage{fancybox}
\usepackage{longtable}
\usepackage{tasks}
\usepackage{pdfpages}

\begin{document}

\title{Challenges, Adaptations, and Fringe Benefits of Conducting Software Engineering Research with Human Participants during the COVID-19 Pandemic}

\titlerunning{Challenges, Adaptations, and Fringe Benefits...}        

\author{Anuradha Madugalla        \and
        Tanjila Kanij \and Rashina Hoda \and Dulaji Hidellaarachchi \and Aastha Pant \and Samia Ferdousi  \and John Grundy 
}


\institute{A. Madugalla\at
              Dept. of Software Systems and Cybersecurity, Monash University, Melbourne, Australia \\
              \email{anu.madugalla@monash.edu}          
           \and
        T. Kanij\at
              Dept. of Software Systems and Cybersecurity, Monash University, Melbourne, Australia \\
              \email{tanjila.kanij@monash.edu}          
           \and
           R.Hoda\at
              Dept. of Software Systems and Cybersecurity, Monash University, Melbourne, Australia \\
              \email{rashina.hoda@monash.edu} 
              \and
            D. Hidellaarachchi\at
              Dept. of Software Systems and Cybersecurity, Monash University, Melbourne, Australia \\
              \email{dulaji.hidellaarachchi@monash.edu}          
           \and
             A. Pant\at
              Dept. of Software Systems and Cybersecurity, Monash University, Melbourne, Australia \\
              \email{aastha.pant@monash.edu}          
           \and
            S. Ferdousi\at
              Dept. of Software Systems and Cybersecurity, Monash University, Melbourne, Australia \\
              \email{samia.ferdousi@monash.edu}          
           \and           
           J. Grundy \at
              Dept. of Software Systems and Cybersecurity, Monash University, Melbourne, Australia \\
              \email{john.grundy@monash.edu}
}

\date{Received: date / Accepted: date}

\maketitle

\begin{abstract}
The COVID-19 pandemic changed the way we live, work and the way we conduct research. With the restrictions of lockdowns and social distancing, various impacts were experienced by many software engineering researchers, especially whose studies depend on human participants. We conducted a mixed methods study to understand the extent of this impact. Through a detailed survey with 89 software engineering researchers working with human participants around the world and a further nine follow-up interviews, we identified the key challenges faced, the adaptations made, and the surprising fringe benefits of conducting research involving human participants during the pandemic. Our findings also revealed that in retrospect, many researchers did not wish to revert to the old ways of conducting human-oriented research. Based on our analysis and insights, we share recommendations on how to conduct remote studies with human participants effectively in an increasingly hybrid world when face-to-face engagement is not possible or where remote participation is preferred. 

\end{abstract}

\section{Introduction}
Working through a global pandemic brought many challenges for professionals around the world. Among them, the COVID-19 pandemic has had a profound impact on many research studies in diverse research domains \cite{omary2020covid}. Necessary measures such as lockdowns, social distancing, and travel restrictions led to various disruptions in research activities \cite{khoo2020lessons} \cite{mourad2020conducting}. Challenges in recruitment, retention of participants, data collection, and access to research sites and facilities have been reported across biomedical, pharmaceutical, data science, clinical, behavioural, social science and other domains. Clinical research was seen to be among the most impacted during the pandemic \cite{tuttle2020impact}. Several ongoing clinical research studies were abandoned or delayed due to the pandemic, leading to significant disruption in the advancement of medical research \cite{riera2021delays}. Further, it was stated that the progress of the studies in areas such as cancer, mental health, and chronic disease were impacted due to the shifting of research funding priorities toward COVID-19 related research \cite{singh2020impact} \cite{tuttle2020impact}. 

As in other research domains, software engineering (SE) practitioners were also seen to be affected by the pandemic. Many research studies have been conducted on the impact of the COVID-19 pandemic on software practitioners, referring to their well-being \cite{juarez2021covid} \cite{ralph2020pandemic}, team behaviour \cite{marinho2021happier}, work-from-home situation \cite{miller2021your} \cite{rodeghero2021please}, and productivity \cite{bezerra2020human}. However, the majority of the studies focused on the impact of the pandemic on SE practitioners, while the impact of the pandemic on SE researchers has not received similar attention. SE researchers were not immune to the challenges presented by the pandemic. While SE researchers in areas such as program analysis, testing, and formal methods would have experienced many professional challenges which should be studied, we were particularly interested in \emph{empirical SE} researchers' experiences. Much of their research relies on human participation, and one would expect this to be majorly impacted in the face of restrictions, such as lockdowns and social distancing, brought on by the pandemic. To understand their experiences, we drafted three overarching research questions:\\

\noindent \textbf{RQ1:} What challenges did empirical SE researchers face during the pandemic?\\
\textbf{RQ2:} How did empirical SE researchers adapt to the challenges?\\
\textbf{RQ3:} Were there any fringe benefits of doing empirical SE research during the pandemic?\\

To answer these questions, we designed a mixed-methods study composed of a survey followed by in-depth interviews with those of the respondents willing to discuss their experiences in further detail. We systematically reached out to 2,190 SE researchers who had collectively published 587 papers in four high-quality research venues for empirical software engineering studies, namely, IEEE Transactions on Software Engineering (TSE), Empirical Software Engineering (EMSE), IEEE International Conference on Software Engineering (ICSE) and Empirical Software Engineering and Measurement (ESEM), with approvals from our Ethics committee and the editors and program co-chairs of these journals and conferences. We were interested in SE studies involving human participants and received 89 responses from relevant SE researchers. We then conducted a further nine in-depth interviews from those willing to share details of their experiences. The mix of primarily qualitative and some quantitative data collected from the survey and interviews was analysed using socio-technical grounded theory for data analysis (STGT4DA) \cite{hoda2021STGT} and descriptive statistical analysis respectively. Further details can be found in the Methodology section.

Based on our analysis, we identified key \textit{challenges} and \textit{adaptations} across research design, recruitment, and data collection. For example, challenges with study environment setup, invitation medium, reduced response rates, handling sensitive data online, and technical challenges were reported. Adaptations included changes to study duration (extending, reducing), leveraging online events, and use of online tools, among others. \textit{Fringe benefits} included improved diversity in recruitment, reduced time and costs, increased flexibility, worldwide research collaborations, and more, presented in the Findings section.

\section{Related Work} 
\subsection{Impact of the Pandemic on Research}

The COVID-19 pandemic produced a significant impact globally for each and every individual, as well as a wide range of domains. In the study \cite{omary2020covid}, Omary et al. discussed the impact of the COVID-19 pandemic on research in two ways -- the impact on research institutions and the impact on researchers. They highlighted that COVID-19 causes new institutional responsibilities and challenges, such as continuing ongoing research, increasing COVID-19-related research, and ensuring safety measurements of the employees and students, whereas researchers faced the challenge of maintaining critical research activities, coping with multiple research approaches, engaging in research remotely, continuously planning and writing research grants, initiating new collaborations and many more. 

Various studies have indicated that clinical research has been the most impacted research due to COVID-19, as observations following patient encounters were hindered due to pandemic restrictions \cite{khoo2020lessons} \cite{mourad2020conducting} \cite{tuttle2020impact}. The studies further highlighted that major health research areas that are unrelated to COVID-19 have also been significantly put on hold or suspended entirely due to a variety of pandemic-related reasons, such as COVID-19-related legal restrictions, logistics, recruitment of vulnerable participants, quality of the collected data and operational concerns \cite{singh2020impact} \cite{villarosa2021conducting}. 

However, these studies also indicated that due to COVID-19, innovative methods and strategies were introduced that have advanced the overall conduct of clinical research.  For example, new approaches were developed to recruit participants and conduct the studies, such as remote visits via telehealth, the use of home-based monitoring technologies, and courier pick-up/ delivery of investigational products to keep the studies going throughout the pandemic. Additionally, risk mitigation has become central to the planning of health research during the pandemic to mitigate the risks faced by healthcare workers and patients due to COVID-19. New policies, safety measurements, ethical considerations and many more in the health research field have been introduced, making clinical research approaches more flexible while maintaining research integrity \cite{singh2020impact} \cite{tuttle2020impact}.

\par The pandemic impacted other research areas, such as social science, economics, education, technology, management etc. and the use of different research methods in these areas. A study conducted by \cite{ruppel2020your} highlighted the challenges faced by social science researchers during the COVID-19 pandemic, specifically referring to the research methods that have been used prior to the pandemic and the need for alternative research strategies. In their study \cite{george2020has}, George et al. discussed the impact of the COVID-19 pandemic on technology and innovation management research and how novel research investigations emerged.  They pointed out that the pandemic declines the interest in large physical infrastructure to make scientific breakthroughs, and virtualization of collaboration has become the key to innovation research. A study conducted by Idnani et al. \cite{idnani2021experience} highlighted the COVID-19 pandemic impact on education, specifically focusing on tuition-dependent institutes in developing countries where online education/ e-learning advancements need to be focused on facing unforeseen situations like COVID-19. 

\par Further, the COVID-19 pandemic majorly impacted on utilising several research methods when conducting research studies. Various studies discussed the challenges and opportunities of participatory research approaches during the COVID-19 pandemic \cite{meskanen2021remote} \cite{hall2021participatory}. They reflect on their experience of conducting remote user studies and highlight the key considerations for successful remote research. In this study, the main focus was ethnographic research practices and pointed out that although the use of a variety of tools was helpful in connecting to the participants, it remained less effective compared to traditional user research. Gruber et al. \cite{gruber2021qualitative} pointed out the importance of adapting research methods to the changing environment due to the COVID-19 pandemic and emphasized the need for researchers to be flexible and creative in addressing the challenges of conducting research with vulnerable populations. 

Focusing on qualitative research projects during the COVID-19 pandemic, Rahman et al. \cite{rahman2021resilient} and Adom et al. \cite{adom2020covid} discussed the ad-hoc adaptations of qualitative research during the pandemic referring to various qualitative research methods and challenges encountered when digitalizing them. Some of the core qualitative data collection methods, namely, interviews, observations, workshops/action research, were considered, and numerous challenges such as distractions from home life, poor internet connections, last minute rescheduling were discussed. A study conducted by \cite{santhosh2021zooming} discusses the use of Zoom as a tool for conducting remote focus groups in the era of social distancing. The study highlighted the advantages and disadvantages of using Zoom for conducting focus groups and provides practical strategies for conducting successful remote focus groups. Irrespective of the challenges faced in adapting a variety of traditional research methods to remote research, the studies such as \cite{renosa2021selfie} complement remote data collection during the COVID-19 pandemic as it increased the accessibility and equity in participant contributions and lower costs. 

\subsection{Impact of the Pandemic on Software Engineering Research}
Studies have been carried out to explore the impact of the COVID-19 pandemic on individuals in various fields, including software engineering\cite{juarez2021covid} \cite{rodeghero2021please}. The majority of these studies have focused on the well-being of software engineers \cite{juarez2021covid} \cite{ralph2020pandemic} \cite{neto2021deep}, software team behaviour \cite{marinho2021happier}, the impact of work from home situation \cite{miller2021your} \cite{rodeghero2021please} \cite{nolan2021work} \cite{butler2021challenges}, and productivity of software development teams \cite{bezerra2020human} \cite{smite2022changes}  during the COVID-19 pandemic. 

For example, in the study  \cite{juarez2021covid}, the developers' well-being during the pandemic was focused as emotions and identified that the majority of the participants expressed to have positive emotions such as happiness, serenity, optimism and etc. In contrast, a study \cite{ralph2020pandemic} conducted on the effect of the pandemic on developers' well-being and productivity indicated that the pandemic has had a negative effect on developers' well-being and productivity. It highlighted the need for greater flexibility and resilience in software development processes. Further, various studies focused on identifying challenges faced by software development teams during the pandemic, including challenges in working from home,  how it impacted communication, collaboration and productivity of the team \cite{nolan2021work} \cite{butler2021challenges}, and the challenges of remote on-boarding of developers ensuring their successful integration to the teams and the organization \cite{rodeghero2021please}. 

However, the majority of these studies have focused on identifying the COVID-19 impact on the software engineering industry and have not discussed their experience in conducting these studies during the pandemic. For example, studies such as \cite{luy2021toolkit} discussed the limitations of conducting user evaluations during the pandemic and the alternative approaches they used to overcome the challenges. In their study, Mendon\c{c}a et al. \cite{de2020dusk} discussed the COVID-19 impact on the R \& D projects where the study mainly focused on development practices than the research aspect. 

\par Although the researchers had to face numerous challenges in continuing their research, some of the adaptations they used are not new to the research. For example, conducting online surveys and online interviews were there even before the COVID-19 pandemic as new opportunities for the conduct of research with the advances in tools and technologies. In their study \cite{archibald2019using}, Archibald et al. discussed the satisfactory level of using Zoom as a data collection method over other interviewing mediums such as face-to-face or telephones.  In another study \cite{ward2015participants}, participants' views on telephone interviews were considered where their overall experience was positive. This shows that some of the adaptations were not new, rather became more common during the pandemic.

\section{Methodology} \label{methodology}

\begin{figure} [b]
  \includegraphics[width=\textwidth]{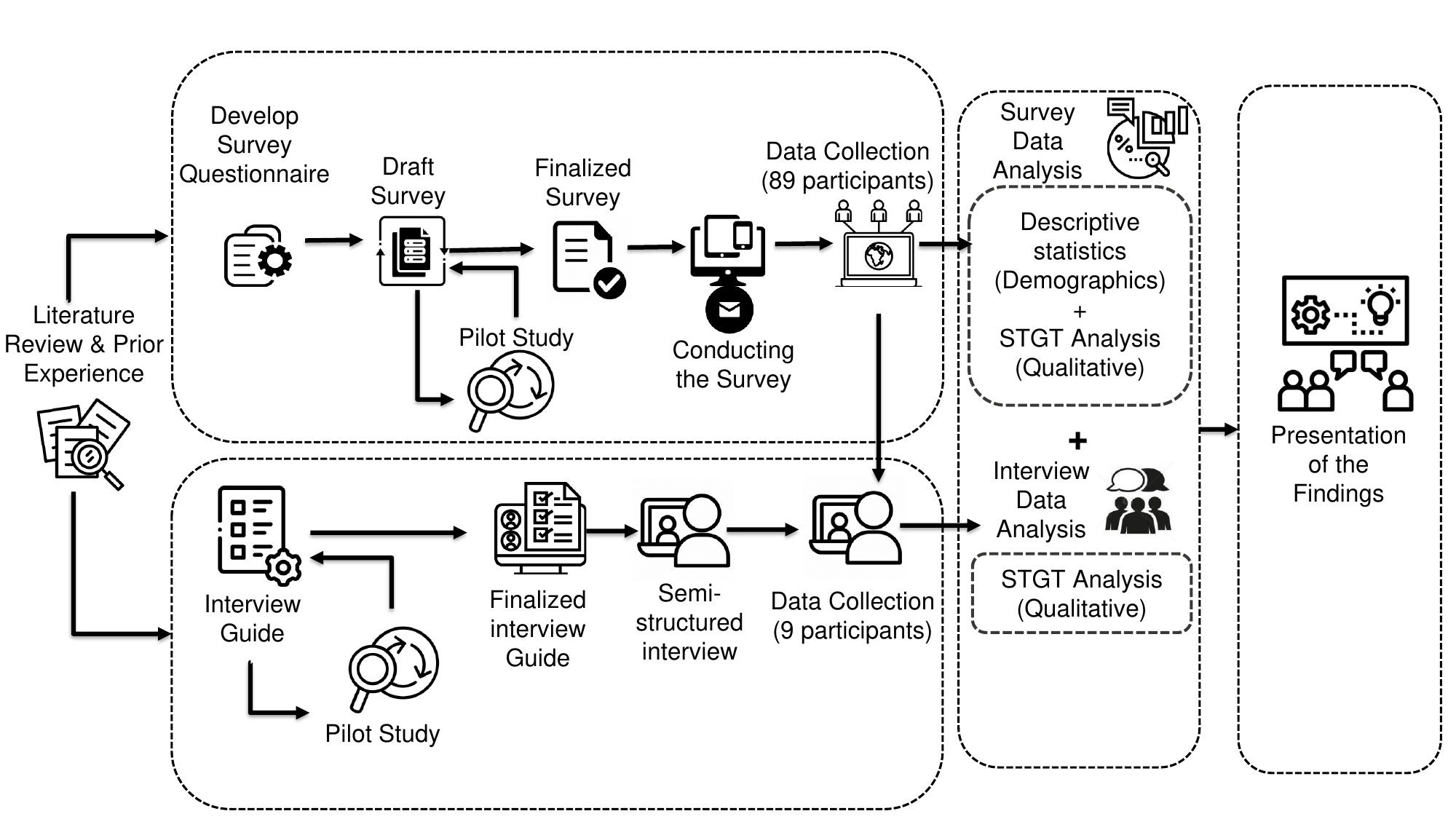}
\caption{\small{Mixed methods research using socio-technical grounded theory for data analysis (STGT4DA) of qualitative data \cite{hoda2021STGT} and statistical analysis of quantitative data. 
}} 
\label{fig:methodology}
\end{figure}

\subsection{Research Design}

Our study was designed as a mixed methods study using a mix of different types of data (primarily qualitative with some quantitative data), and different data collection methods (survey and interviews), using a sequential approach to expand on the findings of the survey with in-depth interviews of willing respondents. Figure. \ref{fig:methodology} summarises the steps followed when designing and carrying out this study.

The questionnaire for the survey was piloted with two initial respondents leading to minor refinements to improve the clarity of the questions and answer choices. Similarly, the semi-structured questions for the interviews were piloted with the first interviewee leading to some refinements in the structure and flow of questions for the remaining eight interviews. We had sought and gained approval from the Human Ethics Committee at Monash University (Reference number: 30921). 

\subsection{Recruitment}
In order to collect quality data from credible sources, we applied a purposive sampling approach to reach out to authors who had papers published in premier Software Engineering conferences and journals, with a special focus on empirical software engineering venues which were likely to publish empirical works involving human participants. We sought approval from the journal editors of the IEEE Transactions on Software Engineering (TSE) and Empirical Software Engineering (EMSE) journals and emailed authors who had their papers published in these journals in the years 2020 and 2021. We also emailed authors who had papers published in the IEEE International Conference on Software Engineering (ICSE) and Empirical Software Engineering and Measurement (ESEM) in the years 2020 and 2021. The aim of this exercise was to find a significant number of relevant authors to survey. Since the impact of the pandemic was different at different times around the world, it was assumed that researchers who had a paper published in 2020 and 2021 likely had experience of conducting research studies involving human participants before the pandemic and had continued doing so during the pandemic, which was validated through the responses of those who agreed to participate. 

Whether a study was conducted with human participants was not always directly discernible from the abstracts. We decided to approach all authors who had papers published in these venues in those two years. We emailed a total of 2,190 researchers who had co-authored 587 papers in the four venues with an invitation to participate in a survey. We explained the purpose of the study in the email and that we were interested in hearing experiences of researchers whose research involved human participants. Participation was anonymous and voluntary. 

We received a total of 89 valid responses from all over the world, representing a 4.1\% response rate. We included all those responses in our analysis. Since the participants were researchers themselves and participated voluntarily without any external incentives, those who were genuinely interested participated, which is reflected in their thorough and meaningful responses. There were no instances of random or less-than-candid responses. Many researchers did not fill out the survey or respond to our email invitation. Some responded informing us that their study did not involve any human participants. We did not include them in the sample. We only included those who responded with a willingness to share experiences from studies involving human participants. Those studies could be either published in that venue (from which we sourced the author) or not yet published or published elsewhere. With this recruitment approach, we were able to reach out to many authors who had relevant experiences to share. In the situation where multiple authors of the same paper wanted to share their experience, we did not filter out any of them as our unit of analysis was individual researchers and their experiences, not individual papers. Multiple people from the same study could have different experiences and perspectives to share, and we wanted to capture these.

\subsection{Data Collection}

Our invitation email contained a direct link to the survey\footnote{\url{https://forms.gle/itWS6pzHTmkE3n4r5}}. The survey was available as a Google Form that took approximately 15-20 minutes to complete. It was structured into four main sections:

\begin{itemize}
    \item Section-A: Four questions on general demographic information. For example, age, gender, years of experience in conducting human-based research in SE and the discipline; 
    \item Section-B: Four questions on research experiences from before the pandemic. For example, the country where the researchers conducted their studies, the number of human-based research studies and the techniques used in data collection prior to the pandemic;
    \item Section-C: Four questions on research experiences during the pandemic. For example, the country where the researchers conducted their studies, the number of human-based research studies and the techniques used in data collection during the pandemic;
    \item Section-D: Seventeen questions on a specific research study (or multiple studies) selected by the respondents that were conducted during the pandemic. For example, the purpose of the study, the participants of the study, and the data collection techniques used for the methods they used such as interviews, surveys, focus groups etc. 
\end{itemize}

While the survey was anonymous, there was an option for respondents to share their contact details if they wanted to participate in a follow-up interview. Based on initial interest registered and later availability, nine researchers proceeded to participate in follow-up semi-structured interviews. None of these nine researchers were co-authors of the same paper. The interviews were conducted over Zoom at a time convenient for the participants and lasted between 30 to 60 minutes. The interview questions focused on one or two specific studies that the researchers had selected to answer Section-D of the survey as well as on the overall experience of conducting SE research with people during the pandemic. In doing so, they identified the specific papers they were referring to and answered questions about the underlying studies, including in-depth examples of challenges encountered, adaptations made and rationales for making those adaptations, as well as any possible benefits of conducting these studies during a pandemic.

\begin{figure} []
  \includegraphics[width=\linewidth]{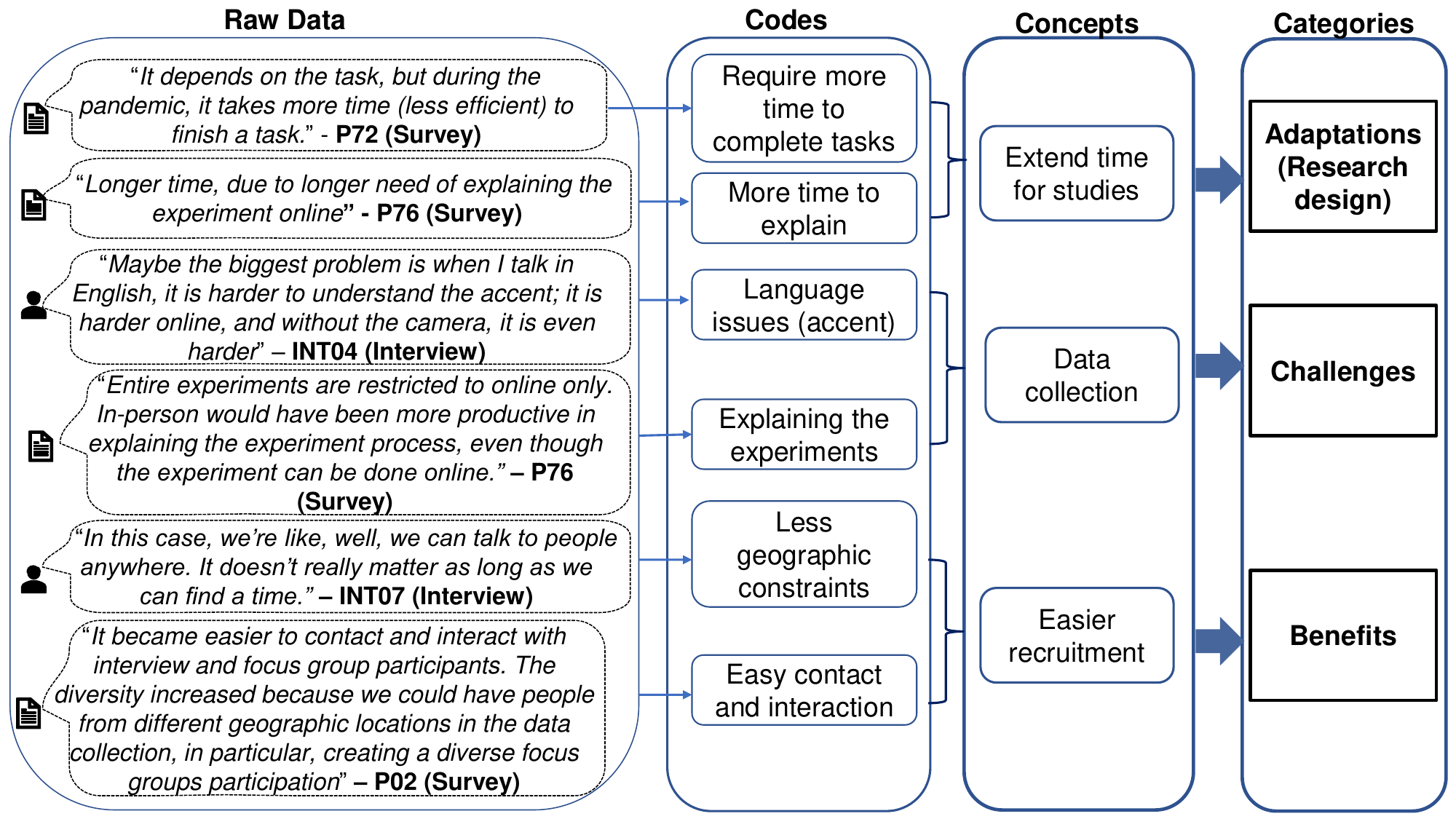}
\caption{\small{Example of applying STGT for data analysis.}}
\label{fig:STGT}
\end{figure}
\begin{figure}
    \centering
    \cornersize{.1} 
    \ovalbox{\begin{minipage}{11cm}
    \textbf{Memo on ``Extending participation time of the research study"}: The researchers had to extend the participation time for various reasons...According to P59, extending the participation time benefited the study.  They considered online participation (via Zoom) as a benefit as it is easier to schedule a meeting time and enabled longer participation times... P76 extended the time to explain their experiment online... On the other hand, P79 mentioned how extending time can negatively impact their research study as participants may quit due to long waiting times.
    \end{minipage}}
    \caption{\small{Example of a memo written as part of the STGT for the data analysis process.}}
    \label{fig:memo}
\end{figure}

\subsection{Data Analysis}

Our survey contained 20 closed-ended questions while the remaining 10 questions were open-ended, where respondents could provide free text responses. As a result, the survey gave rise to both quantitative and qualitative data. The interviews on the other hand gave rise to only qualitative data. The quantitative data was analysed using descriptive statistics and its findings are reported at pertinent points in the Findings section below. 

The qualitative data, from both the open-ended survey questions and the semi-structured interviews, were analysed using socio-technical grounded theory for data analysis (STGT4DA) \cite{hoda2021STGT} which involved socio-technical open coding, constant comparison, and memoing procedures. Given the socio-technical nature of the phenomenon under study in the socio-technical domain of SE and the qualitative data from the surveys and interviews that needed analysing, we found STGT4DA to be well suited to our purpose. Figure \ref{fig:STGT} presents examples of the open coding and constant comparison while Figure \ref{fig:memo} shows one of the memos written during the analysis. The fourth and fifth authors performed the open coding, while the first, second, and third authors reviewed and provided inputs on codes, concepts, and categories and helped refine the findings through critical questioning and feedback. Through discussions, consensus was reached on the most prominent codes, concepts, and categories. Having multiple researchers involved in this way helped improve the richness and strength of the data analysis.

The length of open-ended responses in the survey varied among respondents. Some provided brief responses to the open-ended questions, whereas others elaborated on their experiences. The nine interviews were all in-depth and provided rich examples of challenges, adaptations, and perceived benefits. Using open coding, a variety of codes were generated from both the survey responses and the interviews. All the survey data, from 89 respondents, was analysed together. Since the interviews could be linked to specific publications and were customised accordingly, they were analysed separately from the survey data. The data arising from the nine interviews were also compared to the answers provided in the survey filled by the interviewees ahead of the interview. These findings are reported in the Findings section. The insights gained from the memoing in particular are reported in the Discussion section.

\section{Findings} \label{findings}

In this section, we present the findings of our study, drawn from both the survey responses and the interviews. The first sub-section provides some general information about our researchers and the type of research they conducted, second sub-section presents our findings in response to the three research questions.

\subsection{General Findings}

\subsubsection{Demographics}
Most of the researchers who participated in our study were 31-40 years of age, had between one to six years of experience in conducting software engineering research, and were male. Table \ref{tab:demo} shows the distribution of the survey respondents in terms of gender, age, and experience in conducting software engineering research. While the low number of women may be representative of the wider known gender imbalance across the SE research community, it may also be an indication of the lower number of women publishing during the pandemic. However, we are unable to confirm this from our study. In terms of their geographical distribution, most of our participants were located in Sweden, USA, Brazil while a few represented Oceania and Asia. 

We invited survey respondents to a follow-up interview, we received 9 responses. Table \ref{TABLE: Interview participants' demographics} summarises all the demographic information of the interview participants. 

\begin{table}[h]
\caption{\small Demographics of Survey Respondents}
\label{tab:demo}
\resizebox{\columnwidth}{!}{
\scriptsize
\begin{tabular}{@{}llll@{}}
\toprule
\textbf{Gender} &  & \textbf{Experience} &  \\ \midrule
Men     & \mybarhhigh{35}{70}                         & Less than 1 year      & \mybarhhigh{2}{4}\\
Women   & \mybarhhigh{9}{18}                         & 1 to 3 years         & \mybarhhigh{11}{22}\\
Others   & \mybarhhigh{0.5}{1}                          & 4 to 6 years         & \mybarhhigh{10.5}{21}\\
\cmidrule(r){1-1}
\textbf{Age}   &          & 7 to 9 years         & \mybarhhigh{5}{10}\\
\cmidrule(r){1-1}
20 to 30 years     & \mybarhhigh{8.5}{17}              & 10 to 12 years        & \mybarhhigh{5.5}{11}\\
31 to 40 years     & \mybarhhigh{21}{42}              & 13 to 15 years        & \mybarhhigh{5.5}{11}\\
41 to 50 years     & \mybarhhigh{10}{20}              & 16 to 20 years        & \mybarhhigh{3}{6}\\
51 to 60 years     & \mybarhhigh{3.5}{7}               & 21+ years        & \mybarhhigh{2}{4}\\
60+ years     & \mybarhhigh{1.5}{3}                    & & \\
\bottomrule
\end{tabular}
}
\end{table}

\begin{table*}[h]
\centering
 \caption{\centering \small{Demographics of the Interview Participants}}
\label{TABLE: Interview participants' demographics}
\resizebox{\textwidth}{!}{%
\begin{tabular}{@{}llllllll@{}}
\toprule
\multicolumn{1}{l}{\textbf{\begin{tabular}[c]{@{}l@{}}ID\end{tabular} } }      
& \textbf{Age} & \textbf{Gender} & \textbf{Country}   
& \textbf{\begin{tabular}[c]{@{}l@{}} Job Title \end{tabular}}   
& \textbf{\begin{tabular}[c]{@{}l@{}} Yrs of ex. \end{tabular}}     &\textbf{\begin{tabular}[c]{@{}l@{}} Team \end{tabular}}      
& \textbf{\begin{tabular}[c]{@{}l@{}} Stdy. Dur\end{tabular}}                        
\\ \midrule

INT01 & 36-40 & Male & USA & {\begin{tabular}[c]{@{}l@{}} Assoc. Professor\end{tabular}}  & 4-6 & 5 & \begin{tabular}[c]{@{}l@{}} Approx. 1 \end{tabular}\\
INT02 & 31-35 & Male & Italy & {\begin{tabular}[c]{@{}l@{}} Principal researcher\end{tabular}}  & 4-6 & 3 & 1+\\
INT03 & 26-30 & Female & USA & {\begin{tabular}[c]{@{}l@{}} Principal researcher\end{tabular}}  & 4-6 & 3 & \begin{tabular}[c]{@{}l@{}} Approx. 0.6\end{tabular} \\
INT04 & 31-35 & Male & Brazil & {\begin{tabular}[c]{@{}l@{}} PhD candidate \end{tabular}}  & 1-3 & 5 & \begin{tabular}[c]{@{}l@{}} 3+ \end{tabular} \\

INT05 & 41-45 & Male & Netherlands & {\begin{tabular}[c]{@{}l@{}} Principal researcher \end{tabular}}  & 1-3 & 8 & \begin{tabular}[c]{@{}l@{}} 1+ \end{tabular} \\
INT06 & 31-35 & Male & Iceland & {\begin{tabular}[c]{@{}l@{}} Principal researcher \end{tabular}}  & 7-9 & 3 & \begin{tabular}[c]{@{}l@{}} 1+ \end{tabular} \\
INT07 & 41-45 & Male & USA & {\begin{tabular}[c]{@{}l@{}} Principal researcher \end{tabular}}  & 10-12 &  7 & \begin{tabular}[c]{@{}l@{}} 0.3\end{tabular} \\
INT08 & 36-40 & Male & Thailand & {\begin{tabular}[c]{@{}l@{}} SE researcher \end{tabular}}  & 4-6 & 10 & \begin{tabular}[c]{@{}l@{}} Approx. 1\end{tabular} \\
INT09 & 36-40 & Male & Spain & {\begin{tabular}[c]{@{}l@{}}  SE researcher \end{tabular}}  & 4-6 & 5 & \begin{tabular}[c]{@{}l@{}} Approx. 4 \end{tabular} \\
\bottomrule \hline
\end{tabular}%
}
\end{table*}

\subsubsection{Number of Studies}


When asked about the number of studies conducted per year, we saw a slight reduction in studies during the pandemic , with more researchers opting to conduct less than three studies. Resulting in a 10\% reduction of high achievers who used to conduct more than four studies per year. This is understandable, as they had to face many challenges (Section 4.4). It was interesting to find that there was no significant difference in the number of studies distribution between male and female researchers. Additionally, one might assume that researchers living in countries with less COVID-19 restrictions would have had a lesser impact as against researchers living in other regions. However, we did not find a co-relation between the geographical location of the researcher and the number of studies and this just goes on to show the widespread impact of the pandemic. 

\subsubsection{Data collection methods and participant groups}
We asked about use of six common data collection methods with our researchers : \textit{Interviews, Surveys, Focus groups, Observations, Workshops} and \textit{User evaluations}. Our findings show that each of these methods were adopted to a lesser extent during the pandemic, as shown in Table \ref{tab:methods}. Out of these six methods, observations were the most impacted. This is supported by our findings in Section 4.4, where observations are identified as the most challenging data collection method during the pandemic. In all these methods, during the pandemic online techniques were adopted most as can be expected. Surveys were mostly conducted via online survey platforms while few used email based questionnaires as well. The other five methods: interviews, focus groups, observations, and workshops, all had a significant uptake in the use of video conferencing. Additionally, few used emails and texting app to conduct interviews, few focus groups were conducted via text based social media groups i.e Whatsapp, Messenger and few observations were conducted offline via diary/note taking. Furthermore, both observations and workshops had few adoptions of collaborative tools as well.    

\begin{table}[h]
\setlength\extrarowheight{-5pt}
\caption{\small Reported data collection methods and participant groups}
\label{tab:methods}
\resizebox{\columnwidth}{!}{
\scriptsize
\begin{tabular}{@{}llll@{}}
\toprule
\textbf{Method} &  & \textbf{Group} &  \\ 
\midrule
Interview & \darkgraybar{31}{62} & Sw Developers & \mybarhhigh{31.5}{63} \\
 & \graybar{27} {54} & Sw Managers & \mybarhhigh{14}{28} \\
&  &  \\
Surveys & \darkgraybar{33.5}{67} & Sw Designers & \mybarhhigh{13}{26} \\
& \graybar{30} {60}  & Sw Testers & \mybarhhigh{12}{24} \\
& \\
Focus Grps & \darkgraybar{14.5}{29} & Users & \mybarhhigh{9.5}{19} \\
& \graybar{8.5} {17} &  Academics & \mybarhhigh{7}{14} \\
& \\ 
Observations & \darkgraybar{18}{36} & Others & \mybarhhigh{10}{20}\\
& \graybar{6} {12}\\
& \\
Workshops & \darkgraybar{11}{22} \\
& \graybar{5.5} {11}\\
& \\
User Evals & \darkgraybar{15}{30} \\
& \graybar{10} {20} &\darkgraybar{3}{Pre-Pandemic} \graybar{3} {During-Pandemic}\\
\\
\\
\bottomrule
\end{tabular}
}
\end{table}

When queried about the participant groups they worked with during the pandemic, we found that most have focused on software industry professionals such as software developers, managers, designers, testers and only a few had worked with software users, academics and other groups. The distribution of number of studies conducted with each of these groups is shown in Table \ref{tab:methods}. In terms of the data collection methods, we found that with academics, observations were used slightly more than other methods whereas with other groups all methods were adopted equally. This can be due the fact that as observations take a significant amount of time, academics were happy to allocate those large chunks of time for research as against other groups.

\subsection{Challenges, Adaptations, and Fringe Benefits}


In this section, for ease of reading, we present the challenges, adaptations, and fringe
benefits under three main research steps that were most commonly covered by
the respondents: research design, recruitment, and data collection.

\subsubsection{RQ1: What challenges did SE researchers face during the pandemic?} \label{RQ1}

In response to the first research question, we identified a number of challenges that SE researchers had faced in working on research that involved human participants during the pandemic. These were spread across the research steps of research design, recruitment, and data collection.\\

\noindent \faSearchPlus \hspace{0.1cm} Challenges with Research Design\\
     \par\textbf{Extra effort to re-design}: 
     A number of researchers stated they had to put extra effort into the design for online data collection. This work included ensuring the security of sensitive data (collected online), difficulties with the ethics application process and other challenges with the design. Researchers said they were often exhausted by the extra effort they had to contribute. According to INT07 -\textit{``I was like, this is crazy, we should simplify."}. This is also reflected in discussions of related work where researchers had to re-design studies by adding an additional step to build rapport with interviewees \cite{archibald2019using} or to prepare interviewees to ensure their home backgrounds are not captured \cite{hall2021participatory}. They also had to plan for possible distractions in working from home and address aspects such as interviewees not being used to seeing themselves on screen while keeping them engaged in eliciting in-depth answers \cite{rahman2021resilient}.
     
\textbf{Challenges in the environmental setup}: Along with adjustments to the design, researchers also faced difficulties with setting up the actual data collection environment. The following quote by INT07 is a reflection of this: \textit{``I think the bigger challenge is just getting it set up and going through, I mean, you typically have to install drivers, you have to calibrate, you have to  attach it to your monitor."}. On a few occasions, it was not possible to set the online environment at all, as P33 mentioned \emph{``... we wanted to do an eye-tracking study, but that requires them coming to our lab or us coming to their office to install an eye tracker, and we couldn't do that since we aren't in person".}\\

\noindent \faUserTimes \hspace{0.1cm} Challenges with Recruitment\\

    \textbf{Challenges with invitation medium}:
    Researchers indicated different challenges related to the particular channels they used for sending invitations. According to researchers [INT09, P27], there were fewer replies to \textit{email} invitations. According to some [INT09, P27], recruitment through online conferences was difficult since people tend to multitask during online conference attendance and it becomes difficult to attract their attention. As one said \textit{``People are not 100{\%} focused into the conference...}''. Another challenge with recruitment medium was the loss of some traditional recruitment channels due to the pandemic. A consequence of this loss was reported as getting participants with diverse demographics compared to pre-pandemic. P48 mentioned that \textit{``Usually able to recruit at meetups, but these were not running, so recruiting was mostly online and involved a lot of students rather than professionals."}

    \textbf{Reduced response rate}:
   While this is a common challenge generally faced by all researchers working with participants, it was potentially aggregated with the pandemic. Researchers indicated that the pandemic led to a reduction in their response rate during data collection. There were also uncertainties as to whether participants will join online data collection or not. The following comment by P24 is reflective of this: \textit{``...there was no guarantee participants join on Zoom"}. INT09 also reported that in the absence of in-person invitations, emails were sent via different media which also reduced the response rate, \textit{``During pandemic, you send email to manager and manager send the emails to them, but you get 60-70 {\%} responses"}. Some related work also highlights how some interviewees perceived video interviews as easier to reschedule (particularly last minute), and how this seem to have affected participants’ commitment to attend on time or show up at all \cite{rahman2021resilient}\\

 
\noindent \faFileExcelO \hspace{0.1cm} Challenges with Data Collection\\

       \textbf{Handling of sensitive data online}: Researchers indicated that handling sensitive data during virtual data collection was a real challenge and extra effort was needed. According to INT07 \textit{``We had to build a website that had our tool and pull data from our source code repositories, which is sensitive, and we had to record the responses of the people not anonymised, which was also sensitive."}

        \textbf{Reduced human cues online}: In the absence of in-person data collection, researchers were deprived of the obvious benefits such as observing participants' emotions and body language. INT08 shared - \textit{``...even we don't evaluate their body languages, it's going to be somewhat difficult see how they feel or whether they struggle with the tool''}. This is supported by literature where researchers find having only the neck-up video left out all hand gestures, leg tapping, etc. that may have provided some useful insights \cite{meskanen2021remote}. Additionally, in some of our other work, we also found that groups such as elderly expressed technical concerns in using online methods and instead preferred in-person interviews. Interacting online also made it harder for some to understand the participants. INT04 said \textit{``Maybe the biggest problem is when I talk in English, it is harder to understand the accent, it is harder online, and without the camera, it is even harder''}. But from an EDI perspective some may feel comfortable in switching off their video and responding via typing. This is supported by literature where people with communicative and cognitive disabilities find communication via typing gave them time to think and respond as well as resulting in less anxiety \cite{buchholz2020remote}\\
             
        \textbf{Challenges with diverse data collection methods}: Researchers stated that challenges in data collection depended on the method used:
        
        \begin{itemize}
            \item[$\bullet$] Survey: P63 mentioned that they had to run their surveys for a longer period of time during the pandemic, compared to before.
            \item[$\bullet$] Focus groups: P3 mentioned that finding a common time slot for conducting focus groups was 
 more challenging. They  said, \textit{``Most research guidelines suggest face-to-face interviews and focus groups over virtual and text-based ones. However, finding a time slot with our study participants -especially a common one for the focus groups- became way too difficult when the COVID crisis hit the country."}
           \item[$\bullet$] Observations: P54 reported that online observation of study participants was challenging compared to face-to-face observation. P12 said, \emph{``We wanted to collect observational data of how novices and experts browse stack overflow. Pre-pandemic, we would have brought people into the lab to use an instrumented computer. However, this was not possible, so we screen-recorded remote interactions and used manual annotation which was definitely a bit of a pain.”}
            \item[$\bullet$] Experiments: P76 said online experiments were less productive than face-to-face ones: \textit{``Entire experiments are restricted to online only. In-person would have been more productive in explaining the experiment process, even though the experiment can be done online."} 
            \item[$\bullet$] Interviews, Focus groups: INT07 mentioned that the quality of the data transcriptions was less than the paid transcriptions they used to have before the pandemic. Likewise, a participant [P75] said that building trust with participants during online interviews was challenging: \emph{``Sometimes it is harder to establish the trust when doing online interviews.” -P75}
           \item[$\bullet$] User evaluations: INT07 stated that it was difficult to evaluate models in practice. However, they did not explain the reason for it. Sending equipment to  users to evaluate  models was challenging as the users did not want to keep the equipment, according to INT07. Likewise, INT09 discussed that it was difficult to grab the attention of users for evaluation during the pandemic. INT08 discussed that there could be a difference in the user evaluation settings during the pandemic, which was challenging. 
        \end{itemize}

        \textbf{Technical challenges}:
        Data collection was frequently interrupted due to technical difficulties related to internet connectivity or the facilitation of online sessions. INT08 reported experiencing technical difficulties in facilitating the users to play around with the tool to know about their experience and see the effectiveness of visualisations. Making changes to the technical setup of a tool was also difficult.

        When analysing these challenges based on SE researchers' experience, we found that junior researchers (those with less than 6 years of experience) struggled more in data collection and study advertisement. This may be due to their lack of experience and lack of pre-established connections for advertising venues. Data also showed that senior researchers (more than 6 years) found recruitment and data analysis more difficult than their juniors. This can possibly be attributed to the limited familiarity of senior researchers in using online methods such as social media for recruitment.

\subsubsection{RQ2: How did SE researchers adapt to the challenges?} \label{RQ2}

In response to the second research question, we identified the resilience of SE researchers in continuing with their human-oriented research despite the aforementioned challenges. Respondents share a variety of adaptations they had made to their research design, recruitment techniques, and data collection procedures in the face of the challenges. However, a considerable number of participants mentioned that they did not use any adaptations in any of the stages of their studies (e.g., recruitment, data collection, etc) as they were using similar approaches even before the pandemic. However, they pointed out that they experienced more participants' availability during the pandemic than before. \emph{"We did not make adaptations, but the participants were more available to participate in video calls than to participate in face-to-face discussions" - P20}. \\

\noindent \faHistory \hspace{0.1cm}Adaptations to Research Design\\

    \par \textbf{Extended study duration}: Researchers acknowledged they conducted longer studies and at times it depended on the type of tasks e.g: \emph{``It depends on the task, but during the pandemic it takes more time (less efficient) to finish a task" - P72}. They extended the participation time and even disregarded task completion time or let the participant complete the study without any time limitation, at their own convenience during the pandemic. The study duration was extended,
    \begin{itemize}
        \item[$\bullet$] To explain the study: The researchers had to extend the time of the study as it requires more time to explain the study when conducted in online environment. For example, it was mentioned that when explaining an experiment online takes more time than conducting it face-to-face. \emph{"Longer, the longer time need of explaining the experiment online" -P76}. 
        \item[$\bullet$] To address technical challenges: \emph{"During the pandemic, an extra time is necessary to prepare for possible technical errors in video conferencing" - P28}. For example, extra time was needed to prepare virtual setups, to deal with online connection problems, and blurriness in the video/audio setups.  However, it was also mentioned that sometimes solving technical challenges was beyond researchers. \emph{"Internet connectivity was not stable for most of the participants, and we could not help in any way" - P64}. 
    \end{itemize}

    \textbf{Reduced study duration}: 
    In contrast to the earlier point, some researchers had sought to reduce their study duration.  Reasons  given included:
    \begin{itemize}
        \item[$\bullet$] To keep participants focused: It was mentioned that some researchers were concerned that it maybe difficult to keep participants focused during online interviews/evaluations as most were working remotely and had many work commitments. To make help them stay focused, the study time was reduced in the pandemic. \emph{"We tried to keep participation time shorter in the online setting because it is more difficult to keep people focused during interviews or evaluations" - P14}. 
        \item[$\bullet$] To recruit more participants: Reducing duration of online studies such as interviews, surveys were identified as beneficial in recruiting more participants during pandemic. \emph{"We tried to keep participation time around 15 minutes, helpful in getting more participants" - P22}.
    \end{itemize}

\noindent \faVideoCamera \hspace{0.1cm}Adaptations to Recruitment\\

\par \textbf{Increased use of social media}: 
    Researchers who used to adopt various forms of physical advertisement had to completely move onto online recruitment, especially using social media. \emph{``Before the pandemic, we used many times face-to-face events to advertise our study and get participants, but now we are more limited to online events, social media (e.g., LinkedIn)" -INT09/P54}. Similarly, another participant also mentioned that they used social media like LinkedIn to connect with people and recruit them: \emph{``I usually use LinkedIn, use connection invitations to connect with them." - INT04/P45} The advertising for recruitment was conducted predominantly via social media such as LinkedIn, and Twitter. They used methods such as posting advertisements on these social media, sharing recruitment information via social media groups (e.g. LinkedIn groups), and sending private messages to potential participants (via LinkedIn). For example, an associate professor mentioned that using social media was very effective during the pandemic. \emph{``Because of the pandemic, many have started relying more on social media to connect. Therefore, advertising there has become more effective" - INT01/P35.} 
    
    \textbf{Use of personalized emails}:
    It was pointed out that sending general emails to a wider interested population did not work when recruiting participants for many research studies during the pandemic. Hence,  researchers used personalized emails to potentially interested participants and as a result the response ratio was high. To do this, first, the researchers identified potential participants via their personal networks and then reached out to them using personalized emails rather than sending a general email invitation. INT07, a principal researcher, mentioned that having a unique pitch for each participant is important when recruiting participants via email. \emph{"There is a template that we follow, but we let them know we looked at the change you made last week, we want to ask you about it, making it personalized" - INT07/P33}. 

     \textbf{Leveraging online events}:
    During the pandemic, events such as workshops, talks, conferences, and meetups were conducted fully online. Researchers used these various events to recruit participants for their studies. Online advertisements (posters), online surveys, and Google forms were shared in these events inviting participants. For example, during online talks, interested participants were collected by sharing a survey/Google form. \emph{"Typically we would have a paper sign-up sheet somewhere on campus, but we collected interested participants through a Google Form instead" -P38}. However, recruiting participants from these online events was somewhat challenging as the researchers might not get the target participant groups they want. For example \emph{"usually able to recruit at meetups, but these were not running so recruiting was mostly online and involved a lot of students rather than professionals" - P48}.

    \textbf{Other approaches for recruitment}:
    Researchers said that they used a few other recruitment methods:
    \begin{itemize}
        \item[$\bullet$] Snowballing:  Asking participants to introduce someone, providing incentives to the participants. This helped to build trust, which was quite important in pandemic-led online interactions \cite{wilson2021beyond}
        \item[$\bullet$] Providing attractive incentives to the participants to appreciate the participants' time commitment. This has always been helpful for recruitment in general. During the pandemic, the researchers who worked with developing countries where most had lost their jobs, found this especially helpful \cite{hall2021participatory}.
        \item[$\bullet$] Personal connections: Some chose to meet participants even after the study which helped in reaching out to them for other studies later and was helpful in having contacts of the participants ahead of time.  \emph{"We identified projects and also where we knew we had contacts ahead of time because we have talked to them earlier" -INT07/P33}. Other research has also found the personal touch introduced by referrals and community outreach helpful in recruitment \cite{kim2021lessons}. Some even opted to conduct pre-interview rapport building by having introductory conversations with interviewees before interviews \cite{rahman2021resilient}
        \item[$\bullet$] Involving gatekeepers: Based on related work, some researchers found it hard to build a rapport with participants due to lack of actions such as handshaking and food/drink sharing. To overcome this, some recruitment was conducted via ‘gatekeepers’ who facilitated an introduction or even secured the interview \cite{rahman2021resilient}
        \item[$\bullet$] Convenience sampling:  This was adapted for the studies with limited time. In the pandemic led recruitment it was difficult to build rapport with new entities. Therefore for some studies, convenience sampling was the only way forward \cite{cooksey2022challenges}
    \end{itemize}

\noindent \faFileMovieO \hspace{0.1cm}Adaptations to Data Collection\\

    \par \textbf{Shift to online methods}:  
    All six data collection methods we talked to researchers about (surveys, interviews, focus groups, observations, workshops and user evaluations) were transformed to online approaches. \emph{``interview scripts were revised and minor alterations were introduced to the version used in online interviews.'' - P2}. As a result, these involved audio/video recordings, screen-recording/screen capturing, web-based Q \& A. \emph{``We usually collect only audio recording during and moderator notes, but we adapted those by adding video recording and notes writing on collaborative tools.'' - P21}. Participants moved to online approaches to conduct surveys. P32 said, \emph{``We moved to Google Forms to conduct surveys." - P32.} This shift resulted in benefits for some e.g.: it helped to conduct more interviews within the same day, while it led difficulties to for some e.g.: it was difficult to conduct observations and user evaluation online.      

   \textbf{Better leveraging online tools}: When using various adaptations to data collection methods, the use of online tools played an important role. Whether it was for interactions with participants, observing/capturing participant tasks or as an alternative way for data collection, the researchers used a variety of online tools during the pandemic such as,
    \begin{itemize}
        \item[$\bullet$] Video conferencing tools: Zoom, Skype, and Google Meet were mentioned as the key video conferencing tools that were used to interact with participants in interviews, observations, focus groups and user evaluation studies. \emph{``Before the pandemic, interviews were conducted in-person only for participants residing in the city. During the pandemic, they were given the option of using Zoom as per their convenience'' -P85}. P45 mentioned that they had to use Google Meet to conduct the interview online during the pandemic whereas P47 used MS Teams. 
       \item[$\bullet$] Screen capturing/recording/transcribing: Most video conferencing tools themselves provided facilities for screen capturing, recording and even transcribing (e.g. Zoom). Screen capturing/recording was used in observations and user evaluations to capture participants' tasks/behaviours during the studies and mentioned that they would use it in future studies as it was very successful. \emph{``It was wonderful, I would only do user studies this way going forward'' - INT07/P33}. Similarly, another participant [P59] said, \emph{``I found it easier for subjects to agree to record audio sessions or interviews on video." -P59}
       \item[$\bullet$] New data sources: Some researchers shifted to using alternative ways for data collection, such as using GitHub repository data, or running several iterations with an internal team when they needed to evaluate a tool that they had developed. They also highlighted the importance of finding new approaches to conduct the studies and having a contingency plan to continue the studies. \emph{``I think the COVID situation draws us to find new solutions that we thought we could never use before'' -INT08/P47}. 
    \end{itemize}



\subsubsection{RQ3: Were there any fringe benefits doing research during the pandemic?} \label{RQ3}

Finally, in posing the third question, we were not expecting to find many benefits to be shared. However, we were pleasantly surprised as respondents shared several fringe benefits of the adaptations they had made to their research practices during the pandemic. These are discussed below. 

    \par \textbf{Easier recruitment}: 
    The majority of the researchers mentioned that recruiting participants for their research studies was easier during the pandemic, as there were fewer geographic constraints in recruitment [INT07, INT02, P33]. INT07 mentioned, \textit{ ``In this case, we're like, well, we can talk to people anywhere. It doesn't really matter as long as we can find a time''}. P29 said, \textit{ ``During the pandemic, all interviews were conducted using video conferences. However, it allowed performing interviews with participants overseas.''}. A researcher in the survey [P78] supported this by stating that it was easier to find participants as the participation was online. P78 said,\textit{``It was easier to find participants from several companies, as everyone participated online.''}
    
   \textbf{Expanded participant cohort}: A researcher [INT01] assumed that there might be high chances of getting participants during the recruitment in the pandemic. Another researcher in the survey [P20] reported that the online recruitment process was better than in-person recruitment as it increased the participant number. P20 said, \textit{ ``We did not make adaptations, but the participants were more ``available'' to participate in video calls than to participate in ``in vivo'' discussions.''} However, it is also mentioned that, during the pandemic, getting more companies together was easier, however, harder to get the commitment sometimes. 
   
     \textbf{Increased participant diversity}: 
    A researcher [INT01] highlighted that the diversity of their study participants increased as they were recruited from around the globe. \textit{``The diversity increased because we could have people from different geographic locations in the data collection, in particular, creating diverse focus groups participation'' -[P02]}.  Another participant [P07] quoted: \emph {``There was a flexibility and possibility to reach a larger pool."} -P07
    
    \textbf{Saving time and  budget}: 
    With the shift to an online mode of conducting studies, both researchers and participants could travel less for studies. Some researchers [INT09, INT02] specifically identified less travel time for data collection. INT09 reported, \textit{``Before the pandemic, we have to go to the office and meet''}. Another aspect of time-saving has happened from the perspective of transcription. Researchers in the interview [INT01] and survey [P33, P59] reported that it was easier to transcribe the data and analyse it during the pandemic. A researcher [INT05] mentioned that email-based interviews during the pandemic saved much of their data transcription time. Another researcher [P58] mentioned \textit{``Since data is already digital, ... data analysis is easier since there is less ambiguity and more structure in the data''}. Overall, a considerable amount of researchers participating in the survey indicated that conducting research during the pandemic saved their time in data collection, which eventually saved their research budget and increased individuals' willingness to participate in their studies. \textit{``Reduced travel times for participants or researchers makes planning easier and might increase willingness to participate'' -[P58]}.

     \textbf{Easier data collection}: Some researchers reported that it was easier to collect data during the pandemic [INT01, INT02, INT07, INT09]. Likewise, researchers in the survey [P29] mentioned that they could easily collect the data internationally during the pandemic. Some researchers identified even more benefits to data collection but these were related to specific data collection methods. 
    \begin{itemize}
        \item[$\bullet$] Interviews: A researcher [INT09] mentioned that conducting interviews during the pandemic was easier as recording the interview is much easier. The researcher said, \textit{``In that case, data collection is easier in the pandemic, we can record like what you are doing.''} P33 mentioned the following, \textit{``Since we did interviews and tool evaluation over video conference, it was easy to record and transcribe the sessions''}. A researcher [INT05] also mentioned that email-based interviews during the pandemic provided some time for them to analyse the responses of the participants. INT05 said, \textit{``Email-based interviews provided enough time for us (research team) to process through the responses of the participants.''} 
        \item[$\bullet$] Focus groups: Conducting focus groups online has helped to keep  participants on track with  discussion topics, compared to conducting them in person. \textit{``In in-person focus group, many participants in the same room would tend to go off-topic. However, it was really easy in an online session to keep the discussions on-topic'' - [P30]}.
        \item[$\bullet$] Observations: One of the researchers [INT07] mentioned that virtual observation was easy and asking questions to the participants was also easier during the pandemic. INT07 said, \textit{``If someone turns off the camera, turns off the microphone, they are pretty much invisible, no one sees them. But they can be observed.''}
    \end{itemize} 
    
     \textbf{Accelerated completion time}: 
    INT07 mentioned that it was easier to complete their studies quickly during the pandemic. INT07 said, \textit{``You can get off and hop on another one immediately, so that's you'd call it like a positive adaptation.''} 
    
    \textbf{Increased flexibility in data collection}: INT01 mentioned that the pandemic allowed more flexibility in data collection e.g. it allowed more time for assignments and Zoom meetings. According to a researcher [INT04], conducting the study during the pandemic was easier as there was no negotiation for the interview duration.  
    
     \textbf{Enhanced study flexibility}: 
    Another benefit discussed by a researcher [INT07] was that it was easier to revisit their collected data, know the drawbacks of the data collection method and identify blind spots via users in the studies with human participants during the pandemic. Going back and making changes to the answers of the collected data was also easy, according to some researchers [INT01, INT08]. INT01 mentioned, \textit{``Participants can go back to the software to answer the survey, which was very beneficial.''}
        
    \textbf{Worldwide research collaborations}: 
    The researchers also discussed getting more opportunities for worldwide research collaborations as a benefit. It was mentioned that it was easier for them to approach and collaborate with geographically spread groups with the help of online collaboration technologies. P59 said, \textit{``I believe that the approach of society to online collaboration technologies allows me to collaborate more with subjects with whom I carried out the research. In addition, it seems to me that since there is less socialization, they also accept more initiatives for meetings to discuss professional issues with me''}.

\section{Discussion}

\subsection{Key Findings}
\begin{figure} []
  \includegraphics[width=\linewidth]{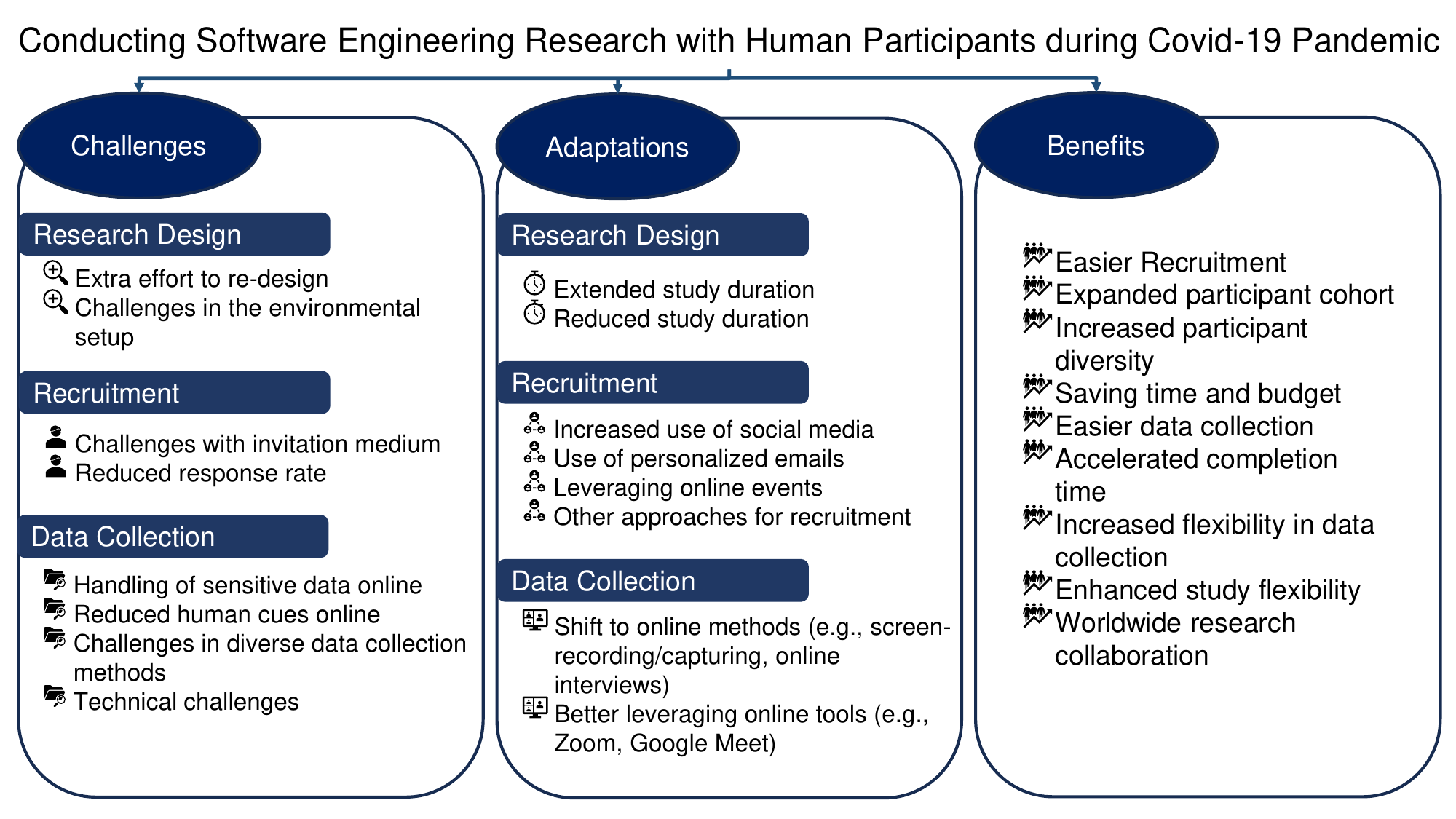}
\caption{\small{Key findings of the study}}
\label{fig: key findings}
\end{figure}

In analysing the data we collected from the survey and  follow-up interviews, we identified a list of challenges SE researchers faced when conducting research with human participants, the adaptations they used as well as several benefits obtained due to the pandemic.
Figure\ref{fig: key findings} summarises these key findings across 
challenges, adaptation and benefits SE researchers reported in our study.

\subsection{Reflections}
As we discussed in our related works section, all research with human participants was impacted by the pandemic. In the field of Medicine, clinical trials were one of the first casualties along with the closing down of laboratory facilities and halt on funding  \cite{chong2010clinical}. There was also a risk of ‘Covidisation’ of academic research, with research grants and output diverted to COVID-19 research in 2020 \cite{riccaboni2022impact}. Sciences such as chemical and biology faced similar challenges, with scientists getting banned from accessing laboratories, shortages of research equipment (e.g. plastic-ware, personal protective equipment), delays in research involving living animals, or requiring human samples \cite{heo2022impacts,bian2020competing}. The most impacted group of scientists were the 'Bench' scientists who relied more on wet labs \cite{gao2021potentially}. Their adaptations included allocating the work-from-home time to plan future experiments and do more reading, which led to more review papers \cite{woolston2021scientists}. Design researchers were another group who were impacted by the pandemic as they needed ``real grass root people and faces" to be the centre of their design practice \cite{RemoteUs20_online}. Similar to SE researchers, they shifted to online interviews and observations. However, as they relied more on paying attention to what people do instead of listening to people, online data collection was more challenging to them \cite{meskanen2021remote}. They adapted their methods to address these challenges by trying to get beyond the question-answer dynamic during interviews. They used ``think-aloud method' and requested proactive commitment from participants by sharing insights with video, pictures, short texts, and audio snippets before the interviews to see more than a bobbing head or a shared screen in a conference call \cite{DoingFie11_online}, \cite{RemoteUs20_online}. As a field of computer science, HCI researchers faced challenges similar to SE researchers with reduced participant engagement in online studies, higher risk of privacy, higher no-show rates and data loss due to technical issues and requiring an extra level of preparation of instructions to send it to participants in advance \cite{shah2021impact}. They also felt recruiting via digital media reduced participant motivation as it narrowed down the persuasive abilities of the research team. Their adaptations included adopting mixed recruitment strategies, balancing simplicity and complexity in studies for better engagement, minding participants’ privacy and proficiency with the adopted tools, and taking advantage of flexible scheduling to increase participant numbers \cite{marques2022hci,van2021remote}.

It can be argued that working remotely had actually started a long time before the pandemic started. During the times of the industrial revolution, most people worked from the office. However, since the 1980s, with the advances in technology, the concept of remote work started emerging, especially with concepts such as ``Global software development" along with ``offshoring" and ``freelancing" \cite{sako2021remote}. Some of the challenges in these remote work were social isolation, missing on training opportunities, work-life balance issues and lack of awareness in teams created by differences in time zones, language and culture \cite{herbsleb2007global,leslie2012flexible}, \cite{charalampous2019systematically}. All these challenges applied to participants SE researchers were interacting with as well.  However, in traditional remote work, the decision to work remotely was a choice; it involved a detailed level of planning, training and preparation. But with the pandemic, none of these were true, and it led to most of the technical challenges that we discussed earlier for SE researchers. Additionally, the physical and mental status of the participants were not similar to remote work teams. For example, during the pandemic, people were not working from remote offices but from bedrooms, kitchen tables and sofas; they had to participate in studies in the middle of distractions from children, partners, siblings and roommates. could have COVID or maybe taking care of ill family members \cite{donnelly2015disrupted}, \cite{ralph2020pandemic}. Therefore, while some of the challenges were similar to traditional remote work, there were a lot of differences led by the enforced nature of pandemic led remote work and lack of preparation.

While we studied SE researchers conducting research with human participants during a pandemic, the irony of our team being one of them is not lost on us. Indeed, the team faced some challenges of our own. These included the usage of terminology, for example ``pre/during'' pandemic, due to the pandemic starting at a different time in different countries. We collected information on which country the researchers were conducting their research in; however, that didn't lead to any interesting findings or correlations. 
Like our respondents we found ourselves adapting our regular practices by providing definitions to terminology such as ``pre/during''. 

Finally, some fringe benefits we experienced include coming up with the recruitment strategy of finding SE researchers from the list of publications in reputed venues. In a pre-pandemic world, we would prefer to contact researchers at conferences. However, all editors were welcoming and supportive of our efforts, by and large, most authors were supportive, too. On one occasion an author expressed disturbance due to receiving multiple invitations resulting from multiple papers at different venues. 
Another observation is that our mixed method approach (survey followed by an interview) enabled us to ask more relevant and customized questions. 
Although participating in a Survey followed by an interview required a significant effort from the participants, those who participated were very forthcoming in their responses. 

In our survey, we have not included a specific section referring to future of hybrid research. However, when we asked about the challenges, adaptations and benefits of doing research during the pandemic, several participants highlighted that the challenges they faced during the pandemic can make a hindrance in keep doing research in a hybrid world in the future. For example, INT02 mentioned that having longer interviews and surveys made difficulties in getting participants online and as a result they are now planning shorter interviews and surveys to use online in future. Further, INT05 highlighted that they would not use slack for focus group studies in future. The reason was that they did not find it useful for longer discussions as it was challenging to follow threads on slack.

\subsection{Threats and Limitations}

We used purposive sampling to find SE researchers with relevant experiences while ensuring credibility and quality. In doing so, we were led by those researchers who had published papers in four prominent SE venues, including two general SE and two empirical SE venues. While it cannot be claimed to be entirely representative, this sampling strategy was successful in providing us with relevant and valuable data and insights for the purpose of the study.

Another potential limitation could be that using Google Forms for data collection may skew the geographical inclusiveness of the sample towards the locations of researchers where Google Forms is easily accessible, and possibly excluding those locations where this is not the case (e.g. China).

While our participants faced several challenges in conducting research during the pandemic, some of these were not limited to pandemic e.g ensuring the security of online collected data. As they faced these challenges during the pandemic they seem to have identified them as being led by pandemic. 

This study focuses on the challenges that were faced during the pandemic. We were not able to gather data on how researchers' experiences changed with post-pandemic as the pandemic was still ongoing when we conducted the study. But analysis that probes more into the comparison between regular remote and pandemic remote work; and the sustainability of the strategies post-pandemic can be explored in future studies. Future studies can also focus on coming up with a set of guidelines for different types of studies (e.g observations) especially in the post-pandemic research.

Since the study focused on the aspects of research related to working with human participants, other steps of research, such as data analysis and writing, were not discussed. However, this does not mean SE researchers may not have had challenges with those research steps. Future studies can take a more holistic look at the SE researcher experience, beyond human-oriented research and into the full research life cycle.

Finally, our own research team faced many of the same challenges reported by participants in conducting this study. Additionally, the time taken to complete the study and write up the findings was delayed in response to being inclusive of one of the research team members being on maternity leave and later on working part-time.

\subsection{Recommendations}
Based on the challenges, adaptations, and fringe benefits reported by SE researchers, we propose the following set of guidelines for SE research with human participants in the post-pandemic world:

\begin{itemize}

    \item[] \faHandORight~\textbf{Plan ahead to minimize technical difficulties}. During online data collection, SE researchers had less control over the environment and sometimes faced technical difficulties. A contingency plan can help reduce the loss in such scenarios. Some technical issues such as internet connectivity are difficult to handle, however, researchers can make a plan for any such scenario and share/discuss it with the participants beforehand, so that the participants know what needs to be done, in case of disruptions (e.g. stop the timer etc).
    
    \item[] \faHandORight~\textbf{Hybrid participation is the new standard}. While recruitment through social media and online events used to be a nice-to-add mechanism in addition to in-person invitation in a pre-pandemic world, we recommend these as standard additions to all recruitment efforts now as it opens up opportunities for more participation and is more inclusive of diverse ways of working and engagement.

    \item[] \faHandORight~\textbf{Making the most of recruitment during online events}. Due to multitasking during online conference attendance, participants often overlook messages. We believe short, attractive graphical messages can help attract their attention while recruiting from online conferences.

    \item[] \faHandORight~\textbf{Leverage features of online tools}. SE researchers discovered features of the online meeting software applications (e.g., recording, screen capture, and auto transcription) to be very helpful for data collection. Researchers can continue using these helpful features.

     \item[] \faHandORight~\textbf{Don't compleyely abandon traditional approaches}. There are still many good reasons to continue traditional approaches to human empirical studies, including in-person interviews, observations and focus groups. Indeed in several of our current studies we are utilising these in preference to online or hybrid studies. This has been done depending on target audience and study e.g. face to face focus groups with elderly participants for one of our studies was by far and away the preferred mode for these participants. In another of our current studies, face to face evaluation of a software tool obviated the need for remote participant challenges with downloading, configuring etc.

\end{itemize}

\section{Conclusion}

The COVID-19 pandemic impacted all professions and domains, Software Engineering research being no different. Our mixed methods study into the experiences of SE researchers from around the world conducting research studies involving human participants during the pandemic revealed several challenges, adaptations, and even some fringe benefits. While some adaptations such as extending study duration and leveraging online events were mostly painful, some fringe benefits such as increased participant diversity and easier data collection were embraced in a way that researchers did not wish to revert to old ways of working. In the post-pandemic world, SE researchers working with human participants can benefit from better study planning and leveraging hybrid modes of participants, effective recruitment from online events, and making the most of the features available on online tools to better support research.

\section{Declarations} 
\subsection{Funding and/or Conflicts of interests/Competing interests:} Conflict of interest include Paul Ralph, Sebastian Baltes, Gianisa Adisaputri, Richard Torkar, Vladimir Kovalenko, Marcos Kalinowski, Nicole Novielli, Shin Yoo, Xavier Devroey, Xin Tan, Minghui Zhou, Burak Turhan, Hideaki Hata, Gregorio Robles, Amin Milani Fard, Rana Alkadhi.

\subsection{Data Availability Statement:} The data sets generated during and/or analysed during the current study are available in a data repository.


%
%

\bibliographystyle{spmpsci}      
\bibliography{References.bib}

\appendix
\section{Appendix : Supplementary Material} \label{A}
A. Madugalla, ‘Supplementary Material for Paper Titled "Challenges, Adaptations, and Fringe Benefits of Conducting Software Engineering Research with Human Participants during the COVID-19 Pandemic"’. Zenodo, Dec. 14, 2023. doi: 10.5281/zenodo.10376868.

\end{document}